# Calculation of quadrupole deformation parameter (β) from reduced transition probability B(E2)↑ for($0_1^+ \to 2_1^+$) transition at (even-even) $^{62\text{-}68}$Zn isotopes

Fatema. H. Obeed[1*] and Ali .K. Hasan[2]

[1,2]Department of Physics, Faculty of Education for Girls, University of Kufa, Al-Najaf, Iraq

[1,*]Corresponding author: fatimahh.alfatlawi@uokufa.edu.iq and alikh.alsinayyid@uokufa.edu.iq[2]

**Abstract**

In this work the excited energy levels, reduced transition probabilities $B(E2)\uparrow$, intrinsic quadrupole moments and deformation parameters have been calculated for $^{62\text{-}68}$Zn isotopes with neutrons (N=32,34,36 and 38). Nushellx code has been applied for all energy states of (fp-shell) nuclei. Shell-model calculations for the zinc isotopes have been carried out with active particles distributed in the $1p_{3/2}$, $0f_{5/2}$ and $1p_{1/2}$ orbits outside doubly-magic closed "$^{56}$Ni" core nucleus. By using (f5p) model space and (f5pvh) interaction, the theoretical results have been obtained and compared with the available experimental results. The excited energies values, electric transition probability B(E2), intrinsic quadrupole moment ($Q_0$) and deformation parameters ($\beta_2$) have been appeared at complete agreement with the experimental values. As well as, the energy levels have been confirmed and determined for the angular momentum and parity of experimental values which have not been well established and determined experimentally. On the other hand, it has been predicted some of the new energy levels and electric transition probabilities for the $^{62\text{-}68}$Zn isotopes under this study which were previously unknown in experimental information.

## 1. Introduction

The structure of the atomic nuclei had witnessed a new contribution in 1948; Maria and Johannes had suggested a nuclear shell model, where the nucleons had been arranged into specific energy states much like the electrons in the atom. Later, much important concept of spin-orbit coupling had been introduced, due to its simplicity to reproduce the shell structure [1]. By the nuclear shell model (SM) there had been attempts to explain the behavior of nuclei in an identical manner such as the Bohr model for atoms, the main concept was to arrange the nucleons which are protons and neutrons, in to a shell structure there major shells that include minor shells called subshells or orbital's[2]. The (SM) of nuclei assumes that nucleons have independent movement in a Hartree-Fock self-consistent potential which basically follows the mass distribution radial dependence, and is necessary to be non-local due to the anti-symmetrization required by the Pauli principle and contains a spin-orbit interaction to produce the correct magic numbers[3]. The nuclei with magic numbers of protons and / or neutrons are not only highly stable but also exhibit some additional characteristics in nuclei which can clearly show that the nuclear shell structure is associated with the independent-particle model for nuclei, each closed-shell configuration in this model gives a convenient first approximation which assume the system under consideration includes a closed-shell core plus valence particles in a valence shell. This approach very successfully explains the ground state properties of nuclei[4]. The (SM) has been greatly included in elucidate all properties of the nuclear levels: excitation energies, electromagnetic transition probabilities and intrinsic quadrupole moments in the medium mass nuclei, particularly study nuclei in the vicinity of doubly magic($^{56}$Ni) nucleus, the study of electromagnetic transition strengths in nuclei can supply available information on the ability of nuclear models to describe many nuclear properties accurately and systematically; this can help in understanding the underlie nuclear structure as well as study the properties of some (even-even) nuclei [5,6]. Nuclear moments have been studied since the beginning of nuclear structure physics, the evaluation of nuclear quadrupole moments is more difficult and challenging than magnetic moment measurements, obviously, to understand the nuclear structure; measurement of the nuclei properties should be made over a large range of isospin or conduct a detailed investigation of some nuclei[7]. Then, the properties of a nucleus with several nucleons outside a closed shell will be described in a first approximation by an inert core (e.g. a doubly magic nucleus such as $^{56}_{28}Ni$ in this present work) with a few nucleons which can move in a certain mixing configuration space and which interact with other nucleons via a residual interaction (particle-particle interactions). The essential inputs in calculations are the model space and effective interaction, which can probe the cogency of the model and parameterizations of the residual interaction via comparison to several experimental parameters (excitation energy, spin/parity, magnetic and quadruple moment), the nuclear moments can be a good check if the parameterization and model space are appropriate, deviations from the model expectations might indicate the presence of configuration that can be mixed into other orbits (without taking into account the chosen model space) or better, parameterized residual interactions[8-10].

## 2. Theory

The necessity of nuclear shell-model calculations inputs are the model-space and effective interactions which represent the most adequate tool for reporting the diagonal matrix of the low energy of Hamiltonian system, the Hamiltonian matrix sizes can be increased significantly with the increase of valence shells number or valence nucleons. Hence, the emphatic truncation of the configuration space is required [11]. Shell-model Hamiltonian in a microscopic approach may be to the shell structure of single particle levels in a spherical potential, the construction start is from a realistic ( N-N) potential by means of many-body perturbation techniques. This approach has long been a central topic of nuclear theory. The solution of the Schrödinger equation for any nuclear system can be given in the following formula [12]:

$$H|\psi_n \rangle = E_n|\psi_n \rangle \qquad (1)$$

Where

$$H = H_o + H_1 \qquad (2)$$

$$H_O = \sum_{i=1}^{A}(T_i + U_i) \qquad (3)$$

$$H_1 = \sum_{i<j}^{A}(V_{ij}^{NN} - \sum_{i=1}^{A} U_i \qquad (4)$$

In order to separate the nuclear Hamiltonian, an one-body potential $U_i$ has been introduced as the sum of a one-body term $H_O$, which describes the independent motion of the nucleons and the interaction $H_1$. Schrodinger equation solutions with $H_O$ are the single nucleon energies (SPE) in a central potential, as observed in single particle (particles or holes) states outside a Doubly-closed shell (DCS) nucleus in its neighbors (DCS ±1). The two body matrix element (TBME) of the residual interaction $H_1$ constitutes the mutual interaction of the valence nucleons on the surface as observed in the DCS ± 2 neighbors' of a magic nucleus, nuclei with only a few particles outside closed shells have also spherical shapes in their ground states, the lowest states in the even-even nuclei are related to the quadrupole vibrations of the nuclear surface; they represent the degrees of freedom, which are easiest to excite, the spherical nuclear shape becomes less and less stable when the number of particles or holes in the unfilled shells increase [13]. the determination of the deviation from spherical balance of the nuclear charge distribution is the electric quadrupole moment ($Q_o$). Therefore electric quadruple moment ($Q_o$) reflects one of the significant quantities to limit the shape of nuclei, the configurations of valence nucleons in an unfilled orbits of nucleus is one of the main causes of its deformation, in another meaning the deformation occurs only when the shells of neutron (N) and proton (Z) are somewhat overcrowded [14]. The intrinsic quadrupole moment ($Q_0$) is defined in the intrinsic frame of reference, It takes three values, zero for nuclei that have a spherically symmetric charge distribution, negative value for oblate nuclei and finally the positive value for prolate nuclei. It is given according to the following formula [15-17]:-

$$Q_0 = \left(\frac{16\pi B(E2)\uparrow}{5e^2}\right)^{1/2} \qquad (5)$$

The symbols in the equation above can be defined as follow :-

($B(E2)\uparrow$ is electromagnetic quadrupole transition probability in Weisskopf unit (W.u.) and $e$ denotes the charge of electron

There is a relationship between B(E2) in unit of $e^2b^2$ and B(E2) in Weisskopf unit (W.u.), this can be according to the following mathematical relationship:-

$$B(E2) \uparrow e^2b^2 = 5.94 \times 10^{-6} \times A^{4/3} \; B(E2) \; W.u. \qquad (6)$$

where $b$ (barn) is the area unit and A is the mass number of nucleus.

The reduced transition probability $B(E2)$ can be defined by:

$$B(E2; i \to f) = \frac{\langle J_f|Q(\lambda)|J_i\rangle^2}{2J_i+1} \qquad (7)$$

Where $\langle J_f|Q(\lambda)|J_i\rangle$ is the reduced matrix element.

B(E2; i → f) depends upon the direction of the transition, for electromagnetic transitions $J_i$ is the higher-energy initial state, the quadrupole transition probability B(E2)↑ is related to the quadrupole deformation parameter $\beta$ of nucleus shape in equilibrium as:-

$$B(E2; 0_1^+ \to 2_1^+) \uparrow = \left(\frac{3}{4}\pi eZR_0^2\right)^2 \beta_2^2 \qquad (8)$$

Where $Z$ is the atomic number and $R_0$ is the average radius of nucleus that is given by:-

$$R_0^2 = 0.0144 \, A^{\frac{2}{3}} \, b \qquad (9)$$

The main aim of the present work is a shell model description of the excited energies, $B(E2\,;0_1^+ \to 2_1^+)\uparrow$ transitions probabilities, intrinsic quadrupole moments, deformation parameters and in a selected set of ( even-even) Zinc isotopes .

## 3. Results and Discussion

In this paper the results is based on nushellx -code for Windows [18]. This code uses is a set of computer codes written by Rae that are used to obtain exact energies, eigenvectors and spectroscopic overlaps for low-lying states in shell model Hamiltonian matrix calculations with very large basis dimensions. It uses a J-coupled proton-neutron basis, and J-scheme matrix dimensions of up to 100 million orders. The formulation comprises f5p-model space consisting of ($lp_{3/2}$, $0f_{5/2}$ and $lp_{1/2}$) shells above the ($^{56}$Ni) nucleus with the (f5pvh) Hamiltonians . (f5pvh) Hamiltonians are Hamiltonians of Van Hienen - Chung and Wildenthal interaction which include the calculations within the fp- shell for( $lp_{3/2}$, $0f_{5/2}$ and $lp_{1/2}$) orbits in code nushellx .

### 3.1 Energy Levels

### 3.1.1 $^{62}$zn Nucleus

Expected shell-model configurations for neutrons and protons in this nucleus primarily involve the ($lp_{3/2}$, $0f_{5/2}$ and $lp_{1/2}$) orbits above the ($^{56}$Ni). From table (1), it can be expected a ($0^+$) ground state for the $^{62}$zn nucleus. The experimental data [19] and our results are shown in table (1). The experimental energies{0.953, 1.804, 2.186,2.341, 2.384,2.743,,2.884 ,3.064, 3.470, 3.707, 3.640 and 5.340} MeV have a good congruence with the expected theoretical values at states{$2_1^+,2_2^+,4_1^+, 0_2^+, 3_1^+, 4_2^+, 2_4^+,2_5^+, 2_8^+,6_1^+,2_{10}^+$ and $0_{10}^+$}, respectively. Theoretically; the energy states{$1_2^+,0_3^+, 2_7^+, 0_4^+, 4_4^+, 2_9^+,3_6^+,3_7^+,5_2^+,3_8^+,3_9^+, 4_8^+,1_9^+,6_3^+ ,0_7^+,5_6^+,7_1^+,0_9^+,6_5^+ ,5_7^+,8_1^+,6_9^+, 7_3^+,8_2^+ ,7_4^+,10_1^+,8_6^+,9_2^+,9_3^+,9_4^+,10_2^+,9_7^+,10_4^+,9_6^+$ and$^+,10_6^+$ } have been confirmed for experimental energy values{3.042,3.160, 3.181,3.310 ,3.374 ,3.590, 3.730, 3.920, 4.021, 4.090, 4.217, 4.380, 4.448, 4.600, 4.620, 4.895, 4.904 ,5.050, 5.131,5.240,5.481, 5.920, 6.081, 6.113, 6.300, 7.500, 7.540, 7.976, 8.300, 8.437, 9.048, 9.465, 9.823, 9.960 and 11.961} MeV for the uncertain practically levels at spins and parities. Experimentally; the energies {4.535, 5.560, 5.700,6.400,6.629, 7.200, 9.800, 10.300 and 10.800} MeV have been predicted in theoretical results with the states{$3_{10}^+,6_7^+ ,5_9^+, ,8_3^+,7_6^+,9_1^+, 9_8^+,9_{10}^+$ and $10_5^+$ } these energies had not been predicted previously at spins and parities. In studied theoretical results, new energy levels have been expected for states; spins and parties such as {2.247;($2_3^+$), 2.668($1_1^+$), 2.907;($4_3^+$),3.035;($2_6^+$), (3.039 to 3.099); for states ($3_2^+$ to $3_4^+$ ),3.398;($5_1^+$),3.433;($1_4^+$), 3.457;($3_4^+$),3.502; ($1_5^+$), 3.540;($4_5^+$), 3.636 ; ($3_5^+$) ,3.651;($0_5^+$),(3.880 to 3.933); for states ($4_6^+$ to $0_6^+$ ), 3.973;($6_2^+$), 4.204; ($1_7^+$), 4.264;($4_7^+$), 4.323;($5_3^+$), 4.374;($1_8^+$),(4.402 to 4.480); for states ($4_9^+$ to $5_4^+$), (4.760 to 4.831); for states ($5_5^+$ to $1_{10}^+$),5.435;($5_8^+$), 5.512;($6_6^+$), (5.674 to 5.835) for states; ($5_{10}^+$ to$7_2^+$ ),6.118;($6_{10}^+$),6.511;($7_5^+$),(6.797 to 7.099) for states;($8_4^+$ to$7_9^+$), 7.320 ;($7_{10}^+$),7.882;($8_7^+$),(8.033 to 8.173) for states ;($8_8^+$ to $8_{10}^+$),8.622;($9_5^+$), 9.347 ; ($9_6^+$) and 9.758;($10_3^+$)}, respectively, these energies and states had not been predicted previously at available experimental information.

### 3.1.2 $^{64}$zn Nucleus

This nucleus contains six nucleons (two protons and four neutrons) outside the $^{56}$Ni core nucleus for configurations spaces (lp$_{3/2}$, 0f$_{5/2}$ and lp$_{1/2}$). In table (2), the experimental energies [20] with states such as{1.799; (2$^+$), 2.793; (2$^+$), ,3.077;(4$^+$) and 4.236;(6$^+$)}MeV , these have a good correspond with the predicted theoretical states and energies{1.832;(2$_2^+$), 2.774;(2$_4^+$),2.927; (2$_5^+$),2.966; (4$_4^+$) and 4.260; (6$_1^+$)}. Theoretically ; it has been expected that the energies {0.938;(2$_1^+$),1.719;(0$_2^+$),2.201; (4$_1^+$),2.378; (0$_3^+$) ,3.061; (1$_3^+$), 3.180; (1$_4^+$),3.524;(4$_7^+$) and 3.843; (5$_2^+$)} are rather in a good agreement with the experimental data {0.991; (2$^+$),1.910; (0$^+$),2.306; (4$^+$),2.609; (0$^+$),3.186; (1$^+$), 3.261; (1$^+$),),3.552; (4$^+$) and 3.852; (5$^+$) } respectively. In the studied calculations, it has been predicted confirmation for the states ;( spins and parities ) as{2$_6^+$,3$_3^+$,3$_4^+$,0$_4^+$,5$_1^+$,2$_8^+$,0$_5^+$,4$_6^+$, 2$_9^+$,1$_5^+$,2$_{10}^+$,0$_6^+$,3$_6^+$, 3$_7^+$,4$_8^+$, 3$_8^+$,4$_8^+$, 3$_9^+$,4$_9^+$ ,0$_7^+$,1$_8^+$,5$_3^+$ ,3$_{10}^+$,4$_{10}^+$,1$_9^+$,1$_{10}^+$ ,5$_5^+$,5$_6^+$,5$_7^+$,6$_6^+$, 5$_9^+$,6$_7^+$,5$_{10}^+$,6$_9^+$, 7$_6^+$,8$_3^+$, 8$_4^+$,8$_7^+$, 8$_{10}^+$, 10$_1^+$,10$_2^+$,9$_6^+$,9$_7^+$,10$_3^+$,11$_1^+$,10$_6^+$,10$_7^+$,10$_{10}^+$,11$_2^+$,} for experimental energies {3.094, 3.196 ,3.205, 3.240,3.285,3.297,3.321,3.415,3.452,3.458, 3.500,3.606 ,3.620, 3.627,3.628 ,3.718 ,3.850, 3.863, 3.880,3.880, 3.889 ,3.932, 3.952 ,4.039 ,4.076, 4.140,4.305,4.420 ,4.467,4.668,4.786, 4.823,5.902, 5.121,5.936, 6.031 ,6.377, 6.998, 7.212,7.556 ,8.181, 8.303,8.4268.580,9.666, 9.804,9.948, 10.460 and 11.626}MeV , respectively. In this nucleus , new energies have been expected in MeV unit with spins and parities as{2.080;(2$_3^+$),(2.488 to 2.768);(4$_2^+$ to 3$_2^+$),2.843; (4$_3^+$) ,3.041 ;(2$_7^{+)}$,3.044;(1$_2^+$), 3.138;(4$_5^+$), 3.540;(1$_6^+$) ,3.868; (6$_2^+$);4.015;(0$_8^+$) ,4.113;( 6$_3^+$) , 6.141; ( 7$_7^+$),6.171;( 7$_8^+$),6.397 ;( 8$_5^+$); 6.471;( 7$_{10}^+$) ,(6.909 to 7.098);( 9$_1^+$ to 8$_9^+$) ,7.685;( 9$_3^+$),8.087;(9$_5^+$),(8.620 to 9.484);( 9$_8^+$ to 10$_5^+$), 10.246;( 10$_8^+$), 10.329;(10$_9^+$). These energies with states are not experimentally known yet. Experimental energies in MeV unit as {3.586,3.698 ,4.154 , 4.181 ,(4.504 to 4.638),4.761 ,5.110 ,(5.337 to5.792 ) ,6.300,6.830,7.380 and 7.900 in the studied calculations specified with states (spins and parities) {1$_6^+$, 6$_1^+$,0$_9^+$,5$_4^+$,(0$_{10}^+$ to 5$_8^+$),7$_1^+$,6$_8^+$,(6$_{10}^+$ to 8$_2^+$),7$_9^+$, 8$_6^+$, 9$_2^+$ and 9$_4^+$} respectively. The experimental value (3.538 MeV ) has been specified in this studied results with state (3$_5^+$) while in experimental data specified with state {2 to 6}.

### 3.1.3 $^{66}$zn Nucleus

This nucleus is described as containing ten nucleons (two protons and eight neutrons) over the close core $^{56}$Ni distribution at (lp$_{3/2}$, 0f$_{5/2}$ and lp$_{1/2}$) orbits . In table (3) , theoretically: agreement has been predicted for energies in MeV unit as {1.022; (2$_1^+$),1.697; ( 2$_2^+$),2.368; (4$_1^+$),2.756; ( 4$_3^+$), 2.979; (0$_3^+$), 2.990; (1$_2^+$),3.050; ( 4$_4^+$), 3.203; (1$_4^+$), 3.239; (2$_8^+$), 3.520; (0$_5^+$) and 4.088, (1$_9^+$)} respectively with experimental energies [21] {1.039,1.872,2.451,2.765, 3.105,3.077 ,3.228, 3.331,3.531 and 4.085} at the same states from (spins and parties ) studied calculations. The theoretical states {3$_1^+$, 0$_4^+$ , 5$_2^+$,4$_8^+$,4$_9^+$,6$_1^+$,5$_4^+$,4$_{10}^+$ ,1$_6^+$,6$_3^+$ ,7$_3^+$, 7$_7^+$, 8$_7^+$ and 9$_2^+$} have been affirmed for experimental energies {2.703 ,3.030,3.709,3.882,3.969 ,4.075, 4.119,4.182,4.223 4.251 5.464,6.292,6.850 and 7.550} . In the studied result the energy value 3.380 MeV has been determined with state (1$_5^+$) while in experimental value 3.432 MeV has been specified with state {1,2$^-$}, as well as theoretical value 3.760;(0$_6^+$) MeV has been determined for energy value ( 3.731;$^+$) in experimental information. Theoretical states {$2_7^+,3_4^+,1_7^+,3_7^+,3_8^+,1_8^+,6_2^+,0_8^+,5_5^+,0_{10}^+,(5_7$ to $7_2^+),8_2^+,7_9^+,8_9^+$ and $10_6^+$} for experimental energies in MeV unit have been determined as {3.241, 3.523 ,3.731, 3.806,3.874,3.924,4.081,4.108, 4.321, 4.454,( 4.527 to 5.352) ,6.000, 6.419,7.170and 11.514} these have not been known previously in the states (spins and parities). The energies and states { 2.227;($2_3^+$),2.536;($4_2^+$),2.777; ($1_1^+$) ,2.869 ;($2_5^+$),2.916; ($3_2^+$),3.004 ;($2_6^+$) ,3.078 ; ($1_3^+$),3.127 ;($3_3^+$), 3.222; ($4_5^+$), 3.350 ; ($5_1^+$) , 3.434;($2_9^+$),3.519; ($4_6^+$); 3.592; ($3_5^+$), 3.614;($1_6^+$) ,3.911;($3_9^+$) ,3.932 ; ($0_7^+$), 4.027 ; ($5_3^+$) ,4.084;($3_{10}^+$), 4.214 ;($0_9^+$), 4.379; ($6_4^+$) ,4.385;($5_6^+$),4.476;($6_5^+$),(5.747 to 5.943) for states ;($7_4^+$ to $7_5^+$), 6.094;($7_6^+$) ,6.164; ($8_3^+$) , 6.343; ($8_4^+$),6.405;($7_8^+$),(6.593 to 7.003) for states ;($8_5^+$ to $8_6^+$),7.161; ($8_8^+$),7.300; ($9_1^+$) and (7.703 to 10.470) for states;( $8_{10}^+$ to $10_5^+$)}have been predicted theoretically, the energies and states have not been previously specified in experimental information .

### 3.1.4 $^{68}$zn Nucleus

In this nucleus valence nucleons{two protons and ten neutrons outside close core $^{56}$Ni distribution at (l$p_{3/2}$, 0$f_{5/2}$ and l$p_{1/2}$) orbits. The predicted excited energy calculations values in MeV unit are in table (4) as follow: The levels {1.126;( $2_1^+$),2.815;( $2_4^+$) and 3.936;( $3_6^+$)} have been very corresponding with experimental excited energy values[22] as{1.077 ;( $2^+$),2.821;( $2^+$) and 3.935; ($3^+$)}. While the levels{1.603 ; ($2_2^+$), 2.489; ( $4_1^+$) and 3.184; ( $0_3^+$)} have been rather agree with experimental excited energy values as {1.883 ;( $2^+$),2.417;( $4^+$) and 3.102;( $0^+$) }. The states{ $4_3^+, 1_2^+, 0_4^+, 2_8^+, 2_9^+, 0_5^+, 4_6^+, 3_7^+, 3_9^+, 4_9^+, 4_{10}^+$ and $1_8^+$} have been affirmed for experimental excited energies{2.959, 3.186, 3.664, 3.709 ,3.942, 3.989,4.027 and 4.284 }. Definite the states {$3_3^+, 1_7^+, 1_9^+$ and $1_{10}^+$} for experimental energies{ 3.496 ;($\underline{3}^+,4^+$),4.414 ;($\underline{1}^+,2^+$), 4.732;($\underline{1}^+,2^+$) and 4.910 ;($\underline{1}^+,2^+$) } MeV. respectively. Assigned the energies with states as {2.675;( $4_2^+$), 2.742;( $2_3^+$), 2.890;( $0_2^+$), 2.980; ( $1_1^+$),3.079; ( $4_4^+$), 3.089; ( $2_5^+$), 3.308; ( $1_3^+$), 3.592; ($2_7^+$), 3.746; ( $3_4^+$), (3.825 to 3.917 ); ($4_5^+$ to $5_2^+$), (4.099 to 4.204); ($1_6^+$ to $5_3^+$), (4.524 to 4.594) ;($3_{10}^+$ to $0_8^+$),4.921; ($0_9^+$) ,5.079;($5_6^+$), 5.530; ($6_4^+$), (5.804 to 8.678); ($6_6^+$ to $7_{10}^+$)}, in the available experimental information, energies and states had not been previously known. These states {$3_1^+, 3_2^+, 3_1^+, 2_6^+, 1_4^+, 1_5^+, 2_{10}^+, 0_6^+, 4_8^+, 0_7^+, 3_7^+, 4_{10}^+, 5_5^+,(0_{10}^+$ to $7_1^+$) and ($5_9^+$ to $6_5^+$)} had been specified for the experimental energies: {2.510,2.955, 3.454, 3.487,3.929,4.061, 4.096, 4.229,4.252, 4.284 ,4.680,4.982, (5.146 to 5.565) and ( 5.610 to 5.990)} which have not been unspecified at the states (spins and parities experimentally).

Table .1 Theoretical and experimental excited levels of $^{62}$Zn nuclei by using (f5pvh) interaction[19]

| Calculations Values | | Experimental Values | | Calculations Values | | Experimental Values | |
|---|---|---|---|---|---|---|---|
| J | Ex(MeV) | J | Ex(MeV) | J | Ex(MeV) | J | Ex(MeV) |
| $0_1^+$ | 0 | $0^+$ | 0 | $5_5^+$ | 4.760 | ------ | ------ |
| $2_1^+$ | 0.870 | $2^+$ | 0.953 | $0_8^+$ | 4.770 | ------ | ------ |
| $2_2^+$ | 1.830 | $2^+$ | 1.804 | $6_4^+$ | 4.827 | ------ | ------ |
| $4_1^+$ | 2.048 | $4^+$ | 2.186 | $1_{10}^+$ | 4.831 | ------ | ------ |
| $0_2^+$ | 2.203 | $0^+$ | 2.341 | $5_6^+$ | 4.942 | $(1^+)$ | 4.895 |
| $2_3^+$ | 2.247 | ------ | ------ | $7_1^+$ | 4.950 | $(7^-)$ | 4.904 |
| $3_1^+$ | 2.312 | $3^+$ | 2.384 | $0_9^+$ | 5.050 | $(2^+)$ | 5.050 |
| $4_2^+$ | 2.553 | $4^+$ | 2.743 | $6_5^+$ | 5.235 | $(6^-)$ | 5.131 |
| $1_1^+$ | 2.668 | ------ | ------ | $5_7^+$ | 5.268 | $(0^+)$ | 5.240 |
| $2_4^+$ | 2.725 | $2^+$ | 2.884 | $0_{10}^+$ | 5.285 | $0^+$ | 5.340 |
| $2_5^+$ | 2.900 | $2^+$ | 3.060 | $8_1^+$ | 5.414 | $(8^+)$ | 5.481 |
| $4_3^+$ | 2.907 | ------ | ------ | $5_8^+$ | 5.435 | ------ | ------ |
| $2_6^+$ | 3.035 | ------ | ------ | $6_6^+$ | 5.512 | ------ | ------ |
| $1_2^+$ | 3.039 | $(0^+)$ | 3.042 | $6_7^+$ | 5.604 | ------ | 5.560 |
| $3_2^+$ | 3.057 | ------ | ------ | $5_9^+$ | 5.632 | ------ | 5.700 |
| $1_3^+$ | 3.076 | ------ | ------ | $5_{10}^+$ | 5.674 | ------ | ------ |
| $3_3^+$ | 3.099 | ------ | ------ | $6_8^+$ | 5.821 | ------ | ------ |
| $0_3^+$ | 3.163 | $(2^+)$ | 3.160 | $7_2^+$ | 5.835 | ------ | ------ |
| $2_7^+$ | 3.231 | $(1^+)$ | 3.181 | $6_9^+$ | 5.952 | $(1^+)$ | 5.920 |
| $0_4^+$ | 3.334 | $(4^+)$ | 3.310 | $7_3^+$ | 6.068 | $(9^-)$ | 6.081 |
| $4_4^+$ | 3.365 | $(1^-)$ | 3.374 | $8_2^+$ | 6.115 | $(8^-)$ | 6.113 |
| $5_1^+$ | 3.398 | ------ | ------ | $6_{10}^+$ | 6.118 | ------ | ------ |
| $1_4^+$ | 3.433 | ------ | ------ | $7_4^+$ | 6.321 | $(8^+)$ | 6.300 |
| $2_8^+$ | 3.444 | $2^+$ | 3.470 | $8_3^+$ | 6.461 | ------ | 6.400 |
| $3_4^+$ | 3.457 | ------ | ------ | $7_5^+$ | 6.511 | ------ | 6.629 |
| $1_5^+$ | 3.502 | ------ | ------ | $7_6^+$ | 6.593 | ------ | ------ |
| $6_1^+$ | 3.504 | $6^+$ | 3.707 | $8_4^+$ | 6.797 | ------ | ------ |
| $4_5^+$ | 3.540 | ------ | ------ | $7_7^+$ | 6.843 | ------ | ------ |
| $2_9^+$ | 3.592 | $(2^+)$ | 3.590 | $8_5^+$ | 6.952 | ------ | ------ |
| $3_5^+$ | 3.636 | ------ | ------ | $7_8^+$ | 7.049 | ------ | ------ |
| $0_5^+$ | 3.651 | ------ | ------ | $7_9^+$ | 7.099 | ------ | ------ |
| $2_{10}^+$ | 3.716 | $2^+$ | 3.640 | $9_1^+$ | 7.239 | ------ | 7.200 |
| $3_6^+$ | 3.845 | $(3^-,4^+)$ | 3.730 | $7_{10}^+$ | 7.320 | ------ | ------ |
| $4_6^+$ | 3.880 | ------ | ------ | $10_1^+$ | 7.366 | $(10^+)$ | 7.500 |
| $1_6^+$ | 3.895 | ------ | ------ | $8_6^+$ | 7.575 | $(8^+)$ | 7.540 |
| $0_6^+$ | 3.933 | ------ | ------ | $8_7^+$ | 7.882 | ------ | ------ |
| $3_7^+$ | 3.950 | $(3^-,4^+)$ | 3.920 | $9_2^+$ | 8.025 | $(9^+)$ | 7.976 |
| $6_2^+$ | 3.973 | ------ | ------ | $8_8^+$ | 8.033 | ------ | ------ |
| $5_2^+$ | 4.005 | $(1^+)$ | 4.021 | $8_9^+$ | 8.088 | ------ | ------ |
| $3_8^+$ | 4.104 | $(4^+)$ | 4.090 | $8_{10}^+$ | 8.173 | ------ | ------ |
| $1_7^+$ | 4.204 | ------ | ------ | $9_3^+$ | 8.379 | $(6^+)$ | 8.300 |
| $3_9^+$ | 4.222 | $(3^-)$ | 4.217 | $9_4^+$ | 8.481 | $(10^+)$ | 8.437 |
| $4_7^+$ | 4.264 | ------ | ------ | $9_5^+$ | 8.622 | ------ | ------ |
| $5_3^+$ | 4.323 | ------ | ------ | $10_2^+$ | 8.874 | $(11^+)$ | 9.048 |
| $4_8^+$ | 4.353 | $(4^+)$ | 4.380 | $9_6^+$ | 9.347 | ------ | ------ |
| $1_8^+$ | 4.374 | ------ | ------ | $9_7^+$ | 9.558 | $(12^+)$ | 9.465 |
| $3_{10}^+$ | 4.399 | ------ | 4.535 | $10_3^+$ | 9.758 | ------ | ------ |
| $4_9^+$ | 4.402 | ------ | ------ | $9_8^+$ | 9.790 | ------ | 9.800 |
| $4_{10}^+$ | 4.472 | ------ | ------ | $10_4^+$ | 9.920 | $(12^+)$ | 9.823 |
| $5_4^+$ | 4.480 | ------ | ------ | $9_9^+$ | 9.992 | $(13^+)$ | 9.960 |
| $1_9^+$ | 4.484 | $(1^+)$ | 4.448 | $9_{10}^+$ | 10.113 | ------ | 10.300 |
| $6_3^+$ | 4.599 | $(7^-)$ | 4.600 | $10_5^+$ | 10.967 | ------ | 10.800 |
| $0_7^+$ | 4.693 | $(0^+)$ | 4.620 | $10_6^+$ | 11.939 | $(16^+)$ | 11.961 |

**Table .2 Theoretical and experimental excited levels of $^{64}$zn nuclei by using (f5pvh) interaction[20]**

| Calculations Values | | Experimental Values | | Calculations Values | | Experimental Values | |
|---|---|---|---|---|---|---|---|
| J | Ex(MeV) | J | Ex(MeV) | J | Ex(MeV) | J | Ex(MeV) |
| $0_1^+$ | 0 | $0^+$ | 0 | $5_6^+$ | 4.335 | $(3,4,5)^+$ | 4.420 |
| $2_1^+$ | 0.938 | $2^+$ | 0.991 | $5_7^+$ | 4.439 | $(0^+)$ | 4.467 |
| $0_2^+$ | 1.719 | $0^+$ | 1.910 | $0_{10}^+$ | 4.497 | -------- | 4.504 |
| $2_2^+$ | 1.832 | $2^+$ | 1.799 | $6_5^+$ | 4.551 | -------- | 4.556 |
| $2_3^+$ | 2.080 | -------- | -------- | $5_8^+$ | 4.659 | -------- | 4.638 |
| $4_1^+$ | 2.201 | $4^+$ | 2.306 | $6_6^+$ | 4.667 | $(6^-)$ | 4.668 |
| $0_3^+$ | 2.378 | $0^+$ | 2.609 | $7_1^+$ | 4.765 | -------- | 4.761 |
| $3_1^+$ | 2.433 | $3^+$ | 2.979 | $5_9^+$ | 4.813 | $(3^+,4^+,5^+)$ | 4.786 |
| $4_2^+$ | 2.488 | -------- | -------- | $6_7^+$ | 4.829 | $(5,6,7)$ | 4.823 |
| $1_1^+$ | 2.712 | -------- | -------- | $5_{10}^+$ | 4.884 | $(3^+,4^+,5^+)$ | 4.902 |
| $3_2^+$ | 2.768 | -------- | -------- | $6_8^+$ | 5.119 | -------- | 5.110 |
| $2_4^+$ | 2.774 | $2^+$ | 2.793 | $6_9^+$ | 5.123 | $(1^+,2^+,3^+)$ | 5.121 |
| $4_3^+$ | 2.843 | -------- | -------- | $6_{10}^+$ | 5.335 | -------- | 5.337 |
| $2_5^+$ | 2.927 | $2^+$ | 3.005 | $7_2^+$ | 5.413 | -------- | 5.413 |
| $4_4^+$ | 2.966 | $4^+$ | 3.077 | $8_1^+$ | 5.526 | -------- | 5.530 |
| $2_6^+$ | 2.972 | $(3)^+$ | 3.094 | $7_3^+$ | 5.565 | -------- | 5.545 |
| $2_7^+$ | 3.041 | -------- | -------- | $7_4^+$ | 5.612 | -------- | 5.613 |
| $1_2^+$ | 3.044 | -------- | -------- | $7_5^+$ | 5.749 | -------- | 5.760 |
| $1_3^+$ | 3.061 | $1^+$ | 3.186 | $8_2^+$ | 5.796 | -------- | 5.792 |
| $3_3^+$ | 3.081 | $(2,3)$ | 3.196 | $7_6^+$ | 5.992 | $(8^+)$ | 5.936 |
| $3_4^+$ | 3.118 | $(3)^+$ | 3.205 | $8_3^+$ | 6.005 | $(8^+)$ | 6.031 |
| $4_5^+$ | 3.138 | -------- | -------- | $7_7^+$ | 6.141 | -------- | -------- |
| $0_4^+$ | 3.141 | $(0)^+$ | 3.240 | $7_8^+$ | 6.171 | -------- | -------- |
| $1_4^+$ | 3.180 | $1$ | 3.261 | $7_9^+$ | 6.289 | -------- | 6.300 |
| $5_1^+$ | 3.210 | $(1^-:5^-)$ | 3.285 | $8_4^+$ | 6.377 | $(9^-)$ | 6.377 |
| $2_8^+$ | 3.253 | $(2)^+$ | 3.297 | $8_5^+$ | 6.397 | -------- | -------- |
| $0_5^+$ | 3.372 | $(1)$ | 3.321 | $7_{10}^+$ | 6.471 | -------- | -------- |
| $4_6^+$ | 3.388 | $(1^-:5^-)$ | 3.415 | $8_6^+$ | 6.789 | -------- | 6.830 |
| $2_9^+$ | 3.391 | $(1, 2^+)$ | 3.452 | $8_7^+$ | 6.857 | $(11^-)$ | 6.998 |
| $1_5^+$ | 3.422 | $(2,3)$ | 3.458 | $9_1^+$ | 6.909 | -------- | -------- |
| $2_{10}^+$ | 3.448 | $(2^+)$ | 3.500 | $8_8^+$ | 6.925 | -------- | -------- |
| $3_5^+$ | 3.514 | $(2\ to\ 6)$ | 3.538 | $8_9^+$ | 7.098 | -------- | -------- |
| $4_7^+$ | 3.524 | $4^+$ | 3.552 | $8_{10}^+$ | 7.202 | $(11^-)$ | 7.212 |
| $1_6^+$ | 3.540 | -------- | 3.586 | $9_2^+$ | 7.424 | -------- | 7.380 |
| $0_6^+$ | 3.600 | $(\leq 4)$ | 3.606 | $9_3^+$ | 7.685 | -------- | -------- |
| $3_6^+$ | 3.613 | $(2\ to\ 6)$ | 3.620 | $10_1^+$ | 7.844 | $(10^-)$ | 7.556 |
| $1_7^+$ | 3.618 | $(0^+,6^-)$ | 3.627 | $9_4^+$ | 7.885 | -------- | 7.900 |
| $3_7^+$ | 3.639 | $(4^+)$ | 3.628 | $10_2^+$ | 8.034 | $(10^-)$ | 8.181 |
| $6_1^+$ | 3.722 | -------- | 3.698 | $9_5^+$ | 8.087 | -------- | -------- |
| $4_8^+$ | 3.742 | $(0^+:4^+)$ | 3.718 | $9_6^+$ | 8.350 | $(12^-)$ | 8.303 |
| $3_8^+$ | 3.842 | $(\leq 3)(\ ^+)$ | 3.850 | $9_7^+$ | 8.427 | $(11^-)$ | 8.426 |
| $5_2^+$ | 3.843 | $5^+$ | 3.853 | $10_3^+$ | 8.531 | $(12^+)$ | 8.580 |
| $3_9^+$ | 3.859 | $(2^+:6^+)$ | 3.863 | $9_8^+$ | 8.620 | -------- | -------- |
| $6_2^+$ | 3.868 | -------- | -------- | $9_9^+$ | 8.733 | -------- | -------- |
| $4_9^+$ | 3.875 | $(0^+:4^+)$ | 3.880 | $9_{10}^+$ | 8.831 | -------- | -------- |
| $0_7^+$ | 3.893 | $(0^+:4^+)$ | 3.880 | $10_4^+$ | 9.250 | -------- | -------- |
| $1_8^+$ | 3.904 | $(2^+,3,4^+)$ | 3.889 | $10_5^+$ | 9.484 | -------- | -------- |
| $0_8^+$ | 4.015 | ------- | ------- | $11_1^+$ | 9.569 | $(14)$ | 9.666 |
| $5_3^+$ | 4.033 | $(4,5)$ | 3.932 | $10_6^+$ | 9.928 | -------- | -------- |
| $3_{10}^+$ | 4.042 | $(3^+,4)$ | 3.952 | $10_7^+$ | 10.073 | $(11^-)$ | 9.804 |
| $4_{10}^+$ | 4.052 | $(0^+:4^+)$ | 4.039 | $10_8^+$ | 10.246 | $(12^-)$ | 9.948 |
| $1_9^+$ | 4.062 | $(5)^+$ | 4.076 | $10_9^+$ | 10.329 | -------- | -------- |
| $1_{10}^+$ | 4.088 | $(1^+,2^+)$ | 4.140 | $10_{10}^+$ | 10.432 | $(13^-)$ | 10.460 |
| $6_3^+$ | 4.113 | -------- | ------- | $11_2^+$ | 11.755 | $(15^-)$ | 11.626 |
| $5_4^+$ | 4.143 | -------- | 4.154 | | | | |
| $0_9^+$ | 4.256 | ------- | 4.181 | | | | |
| $6_4^+$ | 4.260 | $6^+$ | 4.236 | | | | |
| $5_5^+$ | 4.280 | $(1^-,5^-)$ | 4.305 | | | | |

Table .3 Theoretical and experimental excited levels of $^{66}$Zn nuclei by using (f5pvh) interaction[21]

| Calculations Values | | Experimental Values | | Calculations Values | | Experimental Values | |
|---|---|---|---|---|---|---|---|
| J | Ex(MeV) | J | Ex(MeV) | J | Ex(MeV) | J | Ex(MeV) |
| $0_1^+$ | 0 | $0^+$ | 0 | $0_9^+$ | 4.214 | ---- | ---- |
| $2_1^+$ | 1.022 | $2^+$ | 1.039 | $1_{10}^+$ | 4.235 | $(1^-)$ | 4.223 |
| $2_2^+$ | 1.697 | $2^+$ | 1.872 | $6_3^+$ | 4.238 | $(7^-)$ | 4.251 |
| $2_3^+$ | 2.227 | ---- | ----- | $5_5^+$ | 4.352 | ---- | 4.321 |
| $4_1^+$ | 2.368 | $4^+$ | 2.451 | $6_4^+$ | 4.379 | ---- | ---- |
| $3_1^+$ | 2.404 | (3) | 2.703 | $5_6^+$ | 4.385 | ---- | ---- |
| $4_2^+$ | 2.536 | ---- | ----- | $0_{10}^+$ | 4.404 | ---- | 4.454 |
| $4_3^+$ | 2.756 | $4^+$ | 2.765 | $6_5^+$ | 4.476 | ---- | ---- |
| $2_4^+$ | 2.772 | $2^+$ | 2.780 | $5_7^+$ | 4.563 | ---- | 4.527 |
| $1_1^+$ | 2.777 | ---- | ----- | $5_8^+$ | 4.669 | ---- | 4.680 |
| $0_2^+$ | 2.842 | ---- | ----- | $6_6^+$ | 4.752 | ---- | 4.758 |
| $2_5^+$ | 2.869 | $2^+$ | 2.938 | $5_9^+$ | 4.803 | ---- | 4.832 |
| $3_2^+$ | 2.916 | ---- | ----- | $6_7^+$ | 4.872 | ---- | 4.875 |
| $0_3^+$ | 2.979 | $(0^+)$ | 3.030 | $5_{10}^+$ | 4.921 | ---- | 4.918 |
| $1_2^+$ | 2.990 | ---- | ----- | $7_1^+$ | 5.028 | ---- | 5.025 |
| $2_6^+$ | 3.004 | ---- | ----- | $6_8^+$ | 5.133 | ---- | 5.124 |
| $4_4^+$ | 3.050 | $4^+$ | 3.077 | $6_9^+$ | 5.282 | ---- | 5.285 |
| $1_3^+$ | 3.078 | ---- | ----- | $6_{10}^+$ | 5.342 | ---- | 5.331 |
| $3_3^+$ | 3.127 | ---- | ----- | $7_2^+$ | 5.345 | ---- | 5.352 |
| $0_4^+$ | 3.152 | $0^+$ | 3.105 | $7_3^+$ | 5.463 | $(9^-)$ | 5.464 |
| $1_4^+$ | 3.203 | $1^+$ | 3.228 | $7_4^+$ | 5.747 | ---- | ---- |
| $4_5^+$ | 3.222 | ---- | ----- | $8_1^+$ | 5.800 | ---- | ---- |
| $2_7^+$ | 3.231 | ---- | 3.241 | $7_5^+$ | 5.943 | ---- | ---- |
| $2_8^+$ | 3.239 | $2^+$ | 3.331 | $8_2^+$ | 6.086 | ---- | 6.000 |
| $5_1^+$ | 3.350 | ---- | ----- | $7_6^+$ | 6.094 | ---- | ---- |
| $1_5^+$ | 3.380 | $1,2^-$ | 3.432 | $8_3^+$ | 6.164 | ---- | ---- |
| $2_9^+$ | 3.434 | ---- | ----- | $7_7^+$ | 6.263 | $(10^+)$ | 6.292 |
| $4_6^+$ | 3.438 | ---- | ---- | $8_4^+$ | 6.343 | ---- | ---- |
| $3_4^+$ | 3.481 | ---- | 3.523 | $7_8^+$ | 6.405 | ---- | ---- |
| $4_7^+$ | 3.519 | ---- | ---- | $7_9^+$ | 6.453 | ---- | 6.419 |
| $0_5^+$ | 3.520 | $0^+$ | 3.531 | $8_5^+$ | 6.593 | ---- | ---- |
| $3_5^+$ | 3.592 | ---- | ---- | $7_{10}^+$ | 6.597 | ---- | ---- |
| $1_6^+$ | 3.614 | ---- | ---- | $8_6^+$ | 7.003 | ---- | ---- |
| $2_{10}^+$ | 3.621 | $2^+$ | 3.670 | $8_7^+$ | 7.071 | $(8^+)$ | 6.850 |
| $3_6^+$ | 3.657 | $1^-,2^+,3^+$ | 3.689 | $8_8^+$ | 7.161 | ---- | ---- |
| $5_2^+$ | 3.708 | (5) | 3.709 | $8_9^+$ | 7.187 | ---- | 7.170 |
| $1_7^+$ | 3.743 | ---- | 3.731 | $9_1^+$ | 7.300 | ---- | ---- |
| $0_6^+$ | 3.760 | $+$ | 3.738 | $9_2^+$ | 7.560 | $(6^+)$ | 7.550 |
| $3_7^+$ | 3.783 | ---- | 3.806 | $8_{10}^+$ | 7.703 | ---- | ---- |
| $3_8^+$ | 3.868 | ---- | 3.874 | $9_3^+$ | 7.975 | ---- | ---- |
| $4_8^+$ | 3.877 | $(2)^+$ | 3.882 | $10_1^+$ | 8.188 | ---- | ---- |
| $1_8^+$ | 3.903 | ---- | 3.924 | $9_4^+$ | 8.206 | ---- | ---- |
| $3_9^+$ | 3.911 | ---- | ---- | $9_5^+$ | 8.534 | ---- | ---- |
| $0_7^+$ | 3.932 | ---- | ---- | $10_2^+$ | 8.558 | ---- | ---- |
| $4_9^+$ | 3.968 | $(4^+)$ | 3.969 | $9_6^+$ | 8.678 | ---- | ---- |
| $6_1^+$ | 3.968 | $(6^-)$ | 4.075 | $9_7^+$ | 8.927 | ---- | ---- |
| $5_3^+$ | 4.027 | ---- | ---- | $9_8^+$ | 9.206 | ---- | ---- |
| $6_2^+$ | 4.083 | ---- | 4.081 | $9_9^+$ | 9.312 | ---- | ---- |
| $3_{10}^+$ | 4.084 | ---- | ---- | $9_{10}^+$ | 9.358 | ---- | ---- |
| $1_9^+$ | 4.088 | $1^+$ | 4.085 | $10_3^+$ | 9.481 | ---- | ---- |
| $0_8^+$ | 4.113 | ---- | 4.108 | $10_4^+$ | 10.289 | ---- | ---- |
| $5_4^+$ | 4.153 | $(1^-)$ | 4.119 | $10_5^+$ | 10.470 | ---- | ---- |
| $4_{10}^+$ | 4.164 | $(6^+)$ | 4.182 | $10_6^+$ | 11.543 | ---- | 11.514 |

**Table .4 Theoretical and Experimental Excited levels of $^{68}$Zn Nuclei by using (f5pvh) interaction [22]**

| Calculations Values | | Experimental Values | | Calculations Values | | Experimental Values | |
|---|---|---|---|---|---|---|---|
| J | Ex(MeV) | J | Ex(MeV) | J | Ex(MeV) | J | Ex(MeV) |
| $0_1^+$ | 0 | $0^+$ | 0 | $3_8^+$ | 4.335 | ------- | 4.325 |
| $2_1^+$ | 1.126 | $2^+$ | 1.077 | $1_7^+$ | 4.413 | $1^+,2^+$ | 4.414 |
| $2_2^+$ | 1.603 | $2^+$ | 1.883 | $3_9^+$ | 4.460 | $(1^+,2^+)$ | 4.444 |
| $4_1^+$ | 2.489 | $4^+$ | 2.417 | $6_2^+$ | 4.492 | $(1,2^+)$ | 4.496 |
| $3_1^+$ | 2.508 | ------- | 2.510 | $4_9^+$ | 4.501 | $(2^+)$ | 4.512 |
| $4_2^+$ | 2.675 | ------- | ------- | $3_{10}^+$ | 4.524 | ------- | ------- |
| $2_3^+$ | 2.742 | ------- | ------- | $6_3^+$ | 4.532 | ------- | ------- |
| $2_4^+$ | 2.815 | $2^+$ | 2.821 | $5_4^+$ | 4.533 | ------- | ------- |
| $0_2^+$ | 2.890 | ------- | ------- | $0_8^+$ | 4.594 | ------- | ------- |
| $3_2^+$ | 2.938 | ------- | 2.955 | $1_8^+$ | 4.613 | $(1^-)$ | 4.608 |
| $1_1^+$ | 2.980 | ------- | ------- | $4_{10}^+$ | 4.713 | ------- | 4.680 |
| $4_3^+$ | 3.038 | $(4^+)$ | 2.959 | $1_9^+$ | 4.735 | $1,2^+$ | 4.732 |
| $4_4^+$ | 3.079 | ------- | ------- | $1_{10}^+$ | 4.895 | $1,2^+$ | 4.910 |
| $2_5^+$ | 3.089 | ------- | ------- | $0_9^+$ | 4.921 | ------- | ------- |
| $0_3^+$ | 3.184 | $0^+$ | 3.102 | $5_5^+$ | 5.041 | ------- | 4.982 |
| $1_2^+$ | 3.212 | $(1,2^+)$ | 3.186 | $5_6^+$ | 5.079 | ------- | ------- |
| $1_3^+$ | 3.308 | ------- | ------- | $0_{10}^+$ | 5.144 | ------- | 5.146 |
| $2_6^+$ | 3.458 | ------- | 3.451 | $5_7^+$ | 5.193 | ------- | 5.187 |
| $1_4^+$ | 3.480 | ------- | 3.487 | $6_4^+$ | 5.267 | ------- | 5.283 |
| $3_3^+$ | 3.502 | $3^+,4^+$ | 3.496 | $5_8^+$ | 5.304 | ------- | 5.420 |
| $2_7^+$ | 3.592 | ------- | ------- | $7_1^+$ | 5.454 | ------- | 5.565 |
| $0_4^+$ | 3.671 | $(1,2)^+$ | 3.664 | $6_5^+$ | 5.530 | ------- | ------- |
| $2_8^+$ | 3.684 | $(2^+)$ | 3.709 | $5_9^+$ | 5.584 | ------- | 5.610 |
| $3_4^+$ | 3.746 | ------- | ------- | $5_{10}^+$ | 5.612 | ------- | 5.693 |
| $4_5^+$ | 3.825 | ------- | ------- | $6_6^+$ | 5.724 | ------- | ------- |
| $5_1^+$ | 3.845 | ------- | ------- | $6_7^+$ | 5.804 | ------- | ------- |
| $3_5^+$ | 3.917 | ------- | ------- | $6_8^+$ | 5.957 | ------- | ------- |
| $5_2^+$ | 3.917 | ------- | ------- | $7_2^+$ | 5.980 | ------- | ------- |
| $1_5^+$ | 3.921 | ------- | 3.929 | $6_9^+$ | 6.187 | ------- | ------- |
| $3_6^+$ | 3.936 | $3^+$ | 3.935 | $8_1^+$ | 6.270 | ------- | ------- |
| $2_9^+$ | 3.937 | $(7^-)$ | 3.942 | $6_{10}^+$ | 6.411 | ------- | ------- |
| $0_5^+$ | 3.972 | $(1^-,2^+)$ | 4.027 | $8_2^+$ | 6.642 | ------- | ------- |
| $4_6^+$ | 3.983 | $(2^+)$ | 4.061 | $7_3^+$ | 6.652 | ------- | ------- |
| $2_{10}^+$ | 3.991 | ------- | 4.096 | $7_4^+$ | 6.723 | ------- | ------- |
| $0_6^+$ | 4.081 | ------- | 4.102 | $7_5^+$ | 6.944 | ------- | ------- |
| $1_6^+$ | 4.099 | ------- | ------- | $7_6^+$ | 7.103 | ------- | ------- |
| $4_7^+$ | 4.099 | ------- | ------- | $7_7^+$ | 7.615 | ------- | ------- |
| $6_1^+$ | 4.157 | ------- | ------- | $7_8^+$ | 7.986 | ------- | ------- |
| $5_3^+$ | 4.204 | ------- | ------- | $8_3^+$ | 8.171 | ------- | ------- |
| $4_8^+$ | 4.229 | ------- | 4.229 | $7_9^+$ | 8.248 | ------- | ------- |
| $0_7^+$ | 4.255 | ------- | 4.252 | $8_4^+$ | 8.640 | ------- | ------- |
| $3_7^+$ | 4.256 | $(2,3^+)$ | 4.284 | $7_{10}^+$ | 8.678 | ------- | ------- |

## 3.2 Reduced electric quadrupole transition probabilities $B(E2)\uparrow$

In this work the electric quadrupole transition probabilities have been expected for $^{62-68}$Zn nuclei within the framework of the nuclear shell model. Transition levels and up electric quadrupole reduced transition probabilities (B(E2)↑) for the ground state band from {($0^+$ to $2^+$),($2^+$ to $4^+$),($4^+$ to $6^+$),($6^+$ to $8^+$) and ($8^+$ to $10^+$)} of even-even $^{62-68}$Zn isotopes are presented in table(5). The electric quadrupole transition probabilities illustrate a sensitive test for the most modern effective interactions that have been gained to describe fp-shell nuclei. The transition probability has been calculated in this work by using the harmonic oscillator potential (HO, b), where ( b›0 ) for each in-band transition. Core polarization effects have been included by choosing the effective charges for protons {$e_p$= 1.035e , 1.208e,1.140e and 1.102e} but for neutrons {$e_n$= 0.502e ,0.600e, 0.624e and 0.725e} for nuclei $^{62-68}$Zn nuclei respectively . In table (5), the electric quadruple transition probabilities are calculated by using (f5pvh) effective interaction. All of the calculated results for electric quadrupole transition probabilities are reasonably consistent with available experimental data [19-22].

**Table 5. Reduced transition probability $B(E2)$↑ in even $^{62-68}$Zn nuclei[19-22 ]**

| Nuclei | Valence nucleons | $J^+$ | $E_{exp.}$ (MeV) | Spin Sequences | $(B(E2)↑)_{Exp.}$ (W.u) | $(B(E2)↑)_{Cal.}$(W.u) |
|---|---|---|---|---|---|---|
| $^{62}$Zn | 6 | 2<br>4<br>6<br>8<br>10 | 9.538<br>2.186<br>3.707<br>5.481<br>------- | $0_1^+ → 2_1^+$<br>$2_1^+ → 4_1^+$<br>$4_1^+ → 6_1^+$<br>$6_1^+ → 8_1^+$<br>$8_1^+ → 10_1^+$ | 84±(40)<br>46.8±(13-22)<br>27.4±(4)<br>10.330±(26-52)<br>------- | 83.972<br>39.509<br>24. 917<br>21.027<br>14.290 |
| $^{64}$Zn | 8 | 2<br>4<br>6<br>8<br>10 | 0.991<br>2.306<br>3.698<br>5.530<br>------- | $0_1^+ → 2_1^+$<br>$2_1^+ → 4_1^+$<br>$4_1^+ → 6_1^+$<br>$6_1^+ → 8_1^+$<br>$8_1^+ → 10_1^+$ | 100±(30)<br>21.96±(9)<br>--------<br>--------<br>-------- | 100.02<br>47.63<br>32.890<br>22.70<br>5.84 |
| $^{66}$Zn | 10 | 2<br>4<br>6<br>8<br>10 | 1.039<br>2.451<br>4.182<br>-------<br>------- | $0_1^+ → 2_1^+$<br>$2_1^+ → 4_1^+$<br>$4_1^+ → 6_1^+$<br>$6_1^+ → 8_1^+$<br>$8_1^+ → 10_1^+$ | 87.5± (20)<br>32.94<br>21.66± (9)<br>-------<br>------- | 87.481<br>38.46<br>27.97<br>12.45<br>12.07 |
| $^{68}$Zn | 12 | 2<br>4<br>6<br>8<br>10 | 1.077<br>2.417<br>-------<br>-------<br>------- | $0_1^+ → 2_1^+$<br>$2_1^+ → 4_1^+$<br>$4_1^+ → 6_1^+$<br>$6_1^+ → 8_1^+$<br>$8_1^+ → 10_1^+$ | 73.45± (95)<br>19.44± (22)<br>-------<br>-------<br>------- | 73.45<br>24.87<br>19.09<br>9.41<br>------- |

### 3.3.Intrinsic quadrupole moments and deformation parameters

The intrinsic quadrupole moments ($Q_0$) of the $^{62-68}$Zn nuclei have been calculated by eq.(5) for even – even ($^{62-68}$Zn ) isotopes and presented in table (6) which have been compared with the experimental values. The calculated intrinsic quadrupole moments of shell model calculations are in a good agreement with the previous experimental results[23], as well as, the deformation parameters ( of nuclei with proton Z=30 and even neutron (N=30- 38) are obtained by using eq. (8) and presented in table (6). The calculated deformation parameters associated with the frame work of shell model have been compared with the corresponding previous experimental results [23-24].

**Table 6. Intrinsic quadrupole moments and deformation parameters for even $^{62-68}$Zn nuclei[23,24]**

| Nuclei | Spin Sequences | $(B(E2)\uparrow)_{Exp.}$ $(e^2b^2)$ | $(B(E2)\uparrow)_{Cal.}$ $(e^2b^2)$ | $\beta_{2\,(Cal.)}$ | $\beta_{2\,(Exp.)}$[23] | $\beta_{2\,(Exp.)}$[24] | $Q_{0\,(Cal.)}$ (b) | $Q_{0\,(Exp.)}$ (b) |
|---|---|---|---|---|---|---|---|---|
| $^{62}$Zn | $0_1^+ \to 2_1^+$ | 0.1224±58 | 0.122 | 0.216 | 0.216(52) | 0.218(8) | 1.107 | 1.116(41) |
| | $2_1^+ \to 4_1^+$ | 0.0682±51 | 0.0576 | | | | | |
| | $4_1^+ \to 6_1^+$ | 0.0399±22 | 0.0363 | | | | | |
| | $6_1^+ \to 8_1^+$ | 0.0150±29 | 0.0306 | | | | | |
| | $8_1^+ \to 10_1^+$ | -------- | 0.0208 | | | | | |
| $^{64}$Zn | $0_1^+ \to 2_1^+$ | 0.1521±45 | 0.1521 | 0.236 | 0.2342(32) | 0.242(11) | 1.236 | 1.27(6) |
| | $2_1^+ \to 4_1^+$ | 0.0349±13 | 0.0724 | | | | | |
| | $4_1^+ \to 6_1^+$ | -------- | 0.0500 | | | | | |
| | $6_1^+ \to 8_1^+$ | -------- | 0.0345 | | | | | |
| | $8_1^+ \to 10_1^+$ | -------- | $0.88\times10^{-2}$ | | | | | |
| $^{66}$Zn | $0_1^+ \to 2_1^+$ | 0.1386 | 0.1386 | 0.221 | 0.2198(26) | 0.218(8) | 1.180 | 1.164(43) |
| | $2_1^+ \to 4_1^+$ | 0.0522 | 0.0609 | | | | | |
| | $4_1^+ \to 6_1^+$ | 0.0343 | 0.0443 | | | | | |
| | $6_1^+ \to 8_1^+$ | -------- | 0.0197 | | | | | |
| | $8_1^+ \to 10_1^+$ | -------- | 0.0191 | | | | | |
| $^{68}$Zn | $0_1^+ \to 2_1^+$ | 0.1211 | 0.1211 | 0.202 | 0.2015(17) | 0.205(12) | 1.103 | 1.11(7) |
| | $2_1^+ \to 4_1^+$ | 0.0321 | 0.0410 | | | | | |
| | $4_1^+ \to 6_1^+$ | ------- | 0.0314 | | | | | |
| | $6_1^+ \to 8_1^+$ | ------- | 0.0155 | | | | | |
| | $8_1^+ \to 10_1^+$ | ------- | ------- | | | | | |

## 4. Conclusions

In nushellx code, the ( f5pvh) interaction have been performed by using model space ( f5p) for $^{62-68}$Zn nuclei. The predicted low-lying levels (energies, spins, and parities), the reduced probabilities of $(B(E2)\uparrow)$ transitions intrinsic quadrupole moments ($Q_0$) and deformation parameters ($\beta_2$) for even $^{62-68}$Zn nuclei have also been calculated. There is an agreement with some extent between the energy levels that has been calculated for the even-even $^{62-68}$Zn isotopes and the experimental values. The energy levels number has been confirmed and determined for which the angular momentum and parity are not well confirmed and specified empirically. There is an excellent agreement with some energy levels, reduced probabilities of $(B(E2)\uparrow)$ transitions, intrinsic quadrupole moments ($Q_0$) and deformation parameters ($\beta_2$) and the experimental values available for $^{62-68}$Zn isotopes .It has been concluded that the shell model configuration mixing in this region is very successful also new values of energy levels, reduced transition probabilities are predicted in this studied calculations, these values have not been evaluated at the experimental results. This investigation increases the theoretical knowledge of all isotopes with respect to the energy levels and reduced transition probabilities.In this study, the results are very useful for compiling nuclear data table[23,24].

**Data Availability:** The data will be available upon request.

**Conflicts of Interest** The authors have not declared any conflict of interest.